\documentstyle[11pt,newpasp,twoside,epsf]{article}
\begin{document}
\title{Limits on the optical magnitude of PSR 1821-24
 in M28.}
\author{F. K. Sutaria}
\affil{ Inter-University Centre for Astronomy and
Astrophysics, Pune, India.}

\begin{abstract}
The detection of a pulsed X-ray counterpart (RX J1824.2-2R52P) of the 3.05 ms 
pulsar PSR 1821-24, suggests the possibility of  a part of the rotational 
energy loss of this
high spindown rate pulsar being in the optical band.
Archival HST data
for M28 is used here to set upper limits on the optical V-band magnitude of 
PSR 1821-24. The optical limit extends the multiwavelength
observations for this source and  provides a constraint for theoretical 
models of pulsar emission.
\end{abstract}

Pulsed optical and high energy emission from pulsars can be attributed to
either non-thermal acceleration of particles accelerated in the
 pulsar magnetosphere or to emission from the heated polar caps.
Unpulsed optical emission can be caused by the thermal cooling of the 
neutron star. Both pulsed and unpulsed components are indirect probes of 
the poorly known conditions in
the neutron star interior and of the neutron star equation of state (Pavlov 
et al., 1996). 
The isolated radio pulsar PSR1821-24 has a high spin 
down rate leading to rotational energy loss rate of
$2.2 \times 10^{36}$ ergs s$^{-1}$. This source has been detected as
a rotation powered, non-thermal emitter of X-rays, having a 
luminosity $L_x|_{0.1-2.4 KeV} = 
1.6 \times 10^{33}(d/5.1 kpc)^2$ ergs sec$^{-2}$,
by ROSAT HRI observations (Danner et al., 1997 and references therein). 
Since the spindown age of PSR1821-24 is $\simeq 10^7$ years (Table 1), it can be
expected to have a surface temperature  
$T < 10^5$ $\deg$K implying that the thermal emission would be predominantly
in the far-UV to optical bands.  
Few old pulsars ($\tau \sim 10^6$ years) have been detected in the optical.
Thus, an upper limit on the V band magnitude of PSR 1821-24 
would add one more data point to this list, and a confirmed detection would 
constrain the amount of non-thermal optical emission from the neutron 
star magnetosphere.
 
Details of the observation parameters and data analysis are given in 
Sutaria (2000).
Briefly, the data was obtained from a Sept. 1997 HST-WFPC2 exposure of 1120$s$.
The radio position of PSR 1821-24 was obtained from the dense 
radio-timing observation of Cognard et al.(1996).
Accounting for the
proper motion of the pulsar, the co-ordinates of PSR1821-24 relative
to the 1997 epoch are $\alpha = 18^h 24^m 32.0052^s$ and $\delta= -24\deg
52 \arcmin {10 \farcs 7212}$. The absolute astrometric accuracy of WFPC2 
is only $\simeq 0.5$ arcsec, mainly because of  inaccuracies 
of 0.5" to 1" in position of stars used in the Guide Star Catalog.
Taking a lower estimate of 0.5\arcsec uncertainty, at least 8 optical
sources have been detected in this region (Fig. 1). The photometry 
of these sources and their relative offsets has been quoted in Table 2. 
Of the eight candidates within the 0.5\arcsec astrometric "error circle",
stars 4 and 5 are too bright to be considered for pulsar candidates.
In order to determine whether these objects are foreground objects
or not, photometric analysis in another band would be required.
If the optical emission were entirely thermal in nature and hence
strongly age dependent, PSR 1821-24 would be too faint to be detected
in this 1120s exposure (Sutaria, 2000).  
Even assuming that optical component is 
enhanced by non-thermal emission, PSR1821-24 is unlikey to
be as bright as any of the other 6 sources in table 2 because of its
age and distance, thus implying that a visual band magnitude of 23.7 in 
STMAG system is only an upper limit on the optical counterpart for this 
source. 
\begin{figure}
\plotone{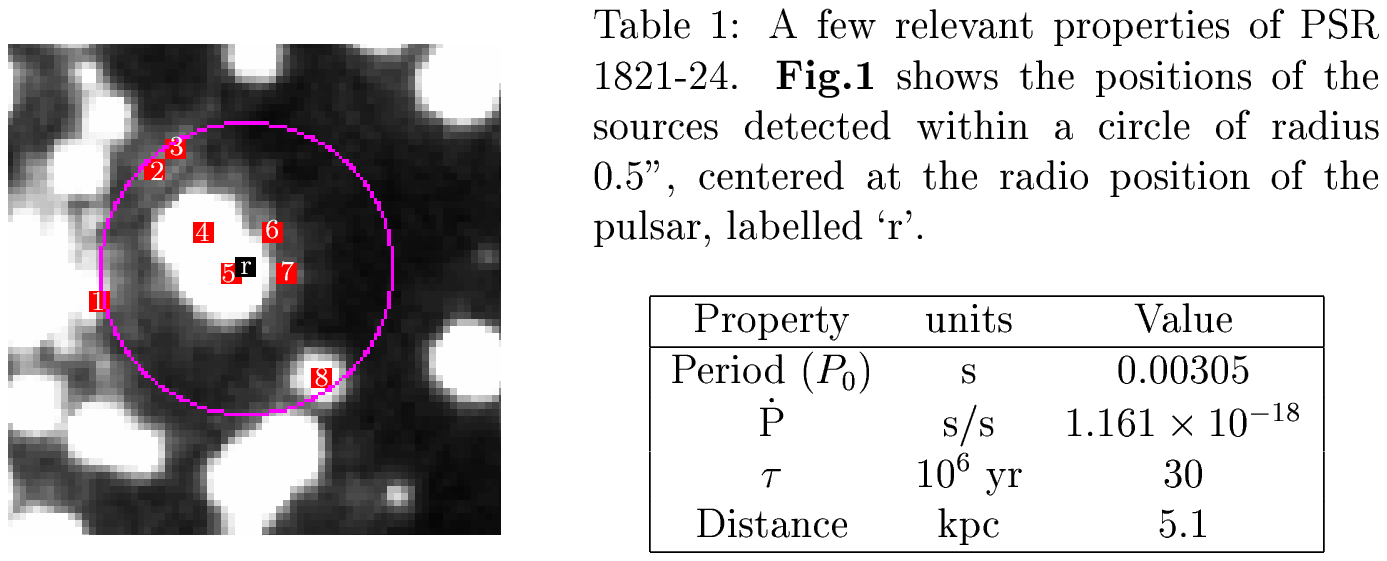}
\end{figure}
\begin{figure}
\plotone{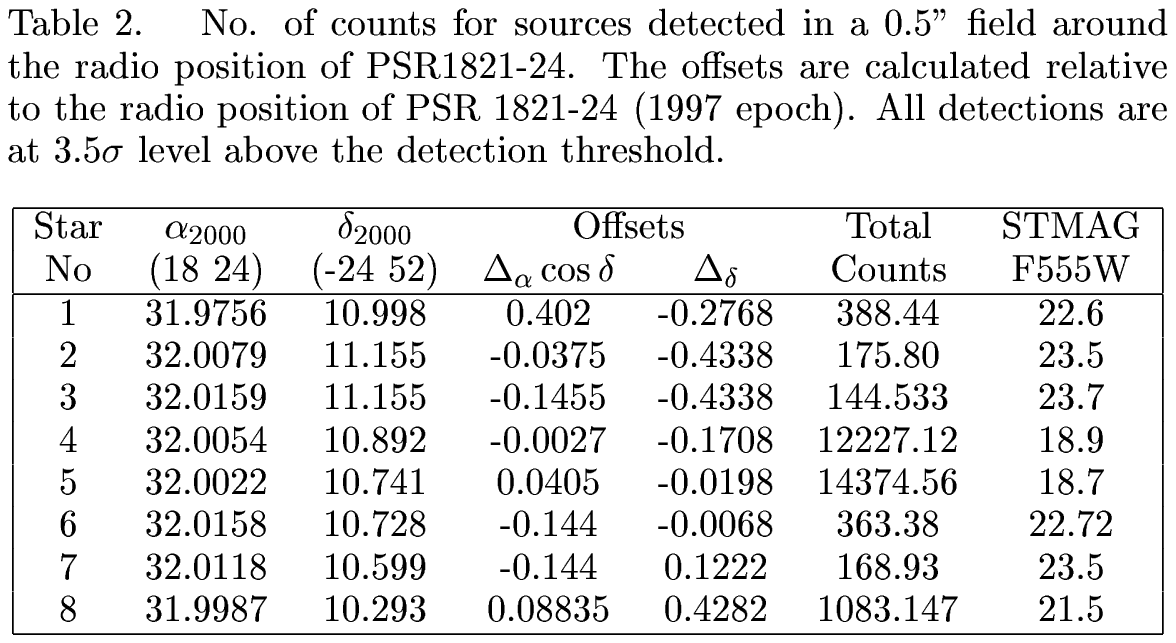}
\end{figure}

\end{document}